\documentclass[useAMS,usenatbib,letter]{mn2e}
\usepackage{psfig} 
\usepackage[dvips]{graphicx}
\usepackage{natbib}
\def\xmm{{\sl XMM-Newton}}
\def\mkn{{Mrk~335}}
\def\tb{T$_{\rm b}$}
\def\hb{H$_\beta$}

\title[Power Spectrum and Time Lags of Mrk 335]{X-ray Variability of the Seyfert 1 Markarian 335: Power Spectrum and Time Lags}  
\author[P. Ar\'evalo et al.]{P. Ar\'evalo$^{1}$\thanks{E-mail: patricia@astro.soton.ac.uk}, I. M. McHardy$^{1}$, D. P. Summons$^{1}$ \\ 
$^1$School of Physics and Astronomy, University of Southampton, Southampton SO17 1BJ, UK\\
}

\begin{document}
\date{Received /Accepted}
\pagerange{\pageref{firstpage}--\pageref{lastpage}} \pubyear{2008}

\maketitle
\label{firstpage}
 
\begin{abstract}
To investigate further the comparison between AGN and black hole X-ray
binaries, we have studied the main X-ray variability properties of the
Seyfert 1 Galaxy \mkn . We put particular emphasis on the X-ray time
lags, which is a potentially important diagnostic of physical
models. From a 100 ksec observation by \xmm\ we show that the power
spectrum of this source is well fitted by a bending power law model,
and the bend time-scale \tb\ is precisely at the value predicted by
the \tb\ vs \hb\ line-width relation of McHardy et al. Variations in
different energy bands show time-scale dependent time lags, where
higher energy bands lag lower ones. The lag, $\tau$, varies as a
function of the Fourier frequency, $f$, of the variability component
in the light curves as $\tau \propto f^{-1}$ at low frequencies, but
there is a sharp cut-off in the lags at a frequency close to the bend
frequency in the power spectrum. Similar behaviour is seen in black
hole X-ray binary systems.  The length of the time lags increases
continuously with energy separation, in an almost log-linear
relation. We show that the lag spectra can be produced by fluctuations
propagating through the accretion flow as long as the energy spectrum
of the X-ray emitting region hardens towards the centre.
\end{abstract}


\section{Introduction}

Understanding the similarities between Active Galactic Nuclei (AGN),
which are powered by accretion onto super-massive black holes, and the
much smaller Galactic black hole X-ray binary systems (BHXRB) is
currently one of the major research topics in high energy
astrophysics. Although some properties might be expected to scale
simply with mass, the fact that the accretion disc is much cooler in
AGN may affect properties which depend on the spectrum of the nuclear
ambient photon field, e.g. any X-ray emission produced by
Comptonisation.

Over recent years considerable attention has been devoted to the
comparison of the X-ray variability properties of AGN and BHXRB and,
in particular, to comparison of their X-ray power spectrum.  It has
been shown that most X-ray bright AGN vary in a manner similar to that
of the persistent BHXRB Cyg~X-1 in the high/soft state (where the
medium energy X-ray spectrum is soft and the flux is high), rather
than in the low/hard state (where the medium energy X-ray spectrum is
hard and the flux is low)
\cite[e.g.][]{psresp,McHardy4051,McHardyMCG,uttleymch05}. In the soft
state, the power density spectrum (PDS) of Cyg~X-1 has a power law
shape of slope -1 at low frequencies (i.e.  $P(f)\propto f^{-\alpha}$
with $\alpha =1$), bending to a steeper slope ($\alpha \geq 2$) above
a bend frequency $f_{b}$, and the PDS of most well observed AGN are
similar. We have shown \citet{mchardynat,kording07} that the
time-scale associated with the bend, $T_{B} = 1/f_{b}$, scales as
black hole mass divided by accretion rate (in Eddington units) from
AGN down to BHXRB. As higher quality AGN X-ray data has become
available, primarily from \xmm , it is now possible to investigate
other X-ray timing measurements which may, ultimately, prove to have
greater diagnostic power than simple bend time-scales, i.e. X-ray time
lags and coherence.

Time lags are normally observed between the fluctuations in different
energy bands, where the hard X-ray bands tend to lag softer ones and
the coherence is the degree to which the hard and soft bands are
correlated. Where the data are good enough, the lags and coherence can
be measured as a function of Fourier frequency. A particularly
important aim, to which observations of lags and coherence can
contribute, is to understand the origin of the X-ray variations in
both BHXRB and AGN. An interesting model for the origin of the
fluctuations was proposed by \cite{lyubarskii}. In this model,
characteristic time-scales of variability, produced by variations in
accretion rate, are associated with each radius in the accretion disc,
with longer time-scales being associated with larger radii, and the
variations propagate inward towards the X-ray emitting region.  This
model provides a natural explanation of the observed linear
relationship between rms variability and flux in both BHXRB and AGN
\citep{rmsflux}.

The propagating fluctuation model has been extended by \citet{kotov01}
and \citet{arevalo} to include an extended X-ray emitting region where
the X-ray spectrum hardens towards the black hole. As the
fluctuations, which might have been produced further out, travel
through this emitting region, they modulated first the emission with a
softer spectrum and later the emission with harder spectra, thereby
naturally giving rise to hard lags. The model has been used to
simultaneously explain both the lags and the X-ray PDS in the high
accretion rate Narrow Line Seyfert 1 (NLS1) Ark~564 \citep{mchardy07}
but it has not yet been widely applied to other AGN. In this paper we
analyse a long \xmm\ observation of the X-ray bright NLS1 \mkn\ and
show how the propagating fluctuation model can simultaneously explain
the lags and the PSD. Using our new measurement of $T_{B}$, we also
investigate how well \mkn\ agrees with the relationship between
$T_{B}$ and the width of the $H_{\beta}$ emission line of
\citep{mchardynat}.

The paper is organised as follows: we briefly discuss the data
reduction in section \ref{data}, calculate the power spectrum and fit
it with a bending power law model in Sec.~\ref{pds}. We study the
relation between different energy bands in Sec.~\ref{coherence} for
the coherence and Sec.~\ref{lags} for the time delays. In
Sec.~\ref{discussion} we summarize the main variability properties of
\mkn\ found in this work.
                           
\section{The data}
\label{data}

\mkn\  was observed by \xmm\ for $\sim 130$ ks in 2006 January 3--5
during revolution 1112. Spectral fitting and spectral variability
analyses of these data have been published by \citet{oneill} and \citet{larsson}.

We used data from the EPIC PN detector \citep{struder}. The PN camera
was operated in Small Window mode, using thin filter. The data were
processed using XMM-SAS v 6.5.0. Source photons were extracted from a
circular region of $5\arcmin$ in radius and a background region of
equal area was chosen on the same chip. Source and background events
were selected by quality flag=0 and patterns=0--4, i.e. only single
and double events were used. The background level was generally low
and stable, except at the beginning and at the end of the observation. We
discarded the first 10 ks and last 12.4 ks to obtain 110.6 ks long
light curves free of background flares.

For the power spectrum and cross spectrum analyses, we constructed
light curves in the 0.2--0.6, 0.6--1, 1--3 and 3--10 keV bands, using
54 sec long time bins, with average count rates of 9.4, 4.5, 4.2 and
0.8 counts/sec, respectively. The bin size was chosen to obtain
$2^{11}$ points in each light curve, which enables us to use the fast
Fourier transform (FFT) in the power spectrum fitting procedure.

\section{power spectrum}
\label{pds}

\begin{figure}
\psfig{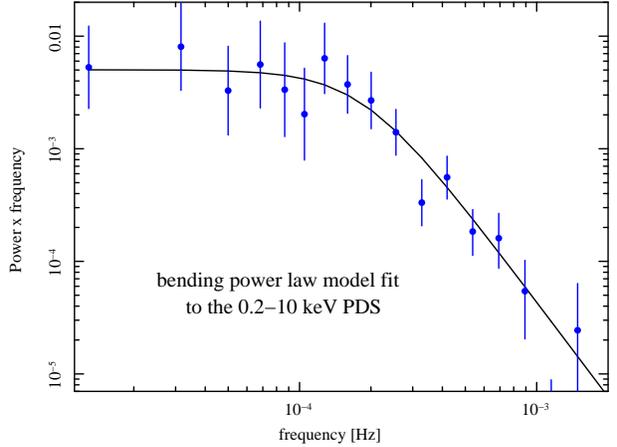}
\caption{0.2--10 keV PDS of \mkn\ plotted in blue dots with error bars, with the best-fitting bending power law model plotted as a solid black line. The power due to Poisson noise, which starts to dominate the variability at $10^{-3}$Hz, has been subtracted.}
\label{hardpsd}
\end{figure}

All the measured AGN power spectra have an approximately $P \propto
1/f$ shape, breaking to a steeper slope at high frequencies
\citep[e.g][]{psresp, markowitz03,McHardyMCG,McHardy4051,markowitz},
similar to the PDS of Cyg~X-1 in the high/soft state. In only one AGN,
Ark~564, an additional, low frequency break has been significantly
detected, below which the PDS levels off to a constant value
\citep{mchardy07}. This shape resembles Cyg~X-1 in either the low/hard
or very high states, which have roughly a doubly-broken power law
shape. In either case, single or double break, the high frequency end
of the PDS has approximately a bending power-law shape with a low frequency slope of -1. For the PDS of \mkn\ we will consider a low
frequency slope of this value to test whether a high frequency bend is
required by the data within the observed frequency band.

We constructed the PDS of \mkn\ light curves through the
discrete Fourier Transform (DFT) \citep{Press}, 
\begin{equation}
P(f_{i})=\frac{2}{\bar x ^2}\frac{\Delta t}{N}|{\rm Re}^2_X(f_{i})+
{\rm Im}^2_X(f_{i})|,
\end{equation}
where ${\rm Re}_X(f_{i})$ and ${\rm Im}_X(f_{i})$ are the real and
imaginary parts of the discrete Fourier transform of the time series
$x(t)$, $\bar x$ is the average count rate, $\Delta t $ is the
sampling time interval and $N$ is the number of points in the light
curve. With this normalisation, the power spectrum is given in terms of
$\sigma^2/\bar x^2$ with units of 1/Hz so the integral of $P(f)$ over
$f$ equals the normalised variance. We binned the periodogram points
in frequency bins, adopting the logarithmic binning method described
by \cite{papadakisbin}. A minimum of 2 points were included in each
frequency bin and the bin width was set at 1.3$\times f_i$, where
$f_i$ is the smallest Fourier frequency in the corresponding bin.

We fitted a bending power law model defined as
\begin{equation}
P(f)=\frac{Af^{-\alpha_L}}{1+(f/f_b)^{{\alpha_H-\alpha_L}}}
\end{equation}
to the PDS of the 0.2--10 keV light curve. The low frequency
slope $ \alpha_L=1$ was fixed and the high frequency slope $\alpha_H$, the
bend frequency $f_b$ and the normalisation $A$ were left as free parameters. We
used the Monte Carlo fitting procedure PSRESP of \citet{psresp}, to
take into account possible distorting effects of red-noise leak and
aliasing and produce robust error bars and goodness of fit criteria.

The parameter space was searched over a grid of multiplicative factor
1.5 in $f_b$, from $10^{-5}$ to $10^{-2.5}$ Hz, and an additive factor
0.2 in $\alpha_H$, from 1 to 4. A finer grid, with steps of 1.1 in
$f_b$ and 0.1 in $\alpha_H$ was used to search over the best fitting
region and the resulting parameter values were essentially the same.

The best-fitting bending power law model, together with the data PDS,
unfolded by observational biases, is shown in Fig.~\ref{hardpsd}. This
model produces a good fit with acceptance probability of 87\%.  The
high frequency slope has a value of $\alpha_H=3.8$ with a 90\%
confidence lower limit of 2.5 and a 99\% limit of 2.2. This slope is
therefore significantly different to the low frequency slope of 1 and
the bend in the PDS is clearly detected. The best-fitting bend
frequency has a value of $1.7_{-1.2}^{+0.85} \times 10^{-4}$Hz, where
the uncertainties correspond to 1$\sigma$ errors. The lower 90\%
confidence limit on the bend frequency, however, falls below $10^{-5}$
Hz, which is the lowest frequency probed. Figure \ref{con_2-10} shows
the confidence contours of this fit over the entire parameter space
explored. As shown below, the bend in the PDS appears to shift to
higher frequencies for higher energy bands. The 99\% confidence lower
limit on the bend frequency falls inside the available parameter space
for the 1--3 and 3--10 keV bands. Figure \ref{con1-3} shows the
confidence contours of the fit to the 1--3 keV band PDS, where the
bend frequency is well bounded. 

Fitting the 0.2-10 keV band PDS, allowing the low frequency slope
$\alpha_l$ to vary produces essentially the same values:
$\alpha_H=3.8_{-1}^{+**}$, $f_b=1.6_{-1.2}^{+1.6} \times 10^{-4}$ Hz ,
and a best fitting value of $\alpha_l=1.1$, but no useful constraints
can be placed on $\alpha_l$.

\begin{figure}
\psfig{file=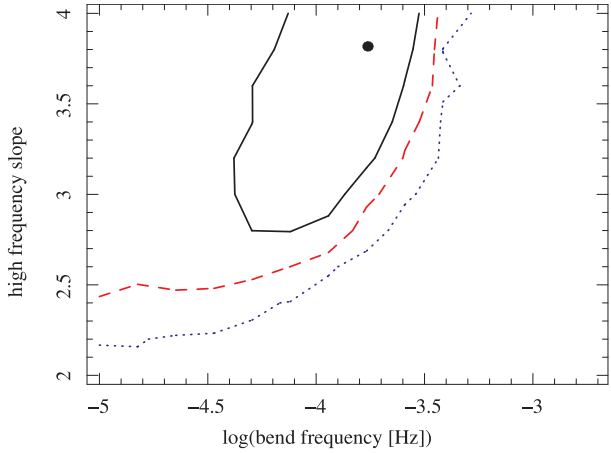,angle=270,width=8.0cm}
 \caption{\label{con_2-10}Confidence contours of the bending power law fit to the 0.2--10 keV PDS. The 68\%, 90\% and 99\% levels are shown by the solid, dashed and dotted lines, respectively. The best fitting values are marked by the black dot.}
\end{figure}

\begin{figure}
\psfig{file=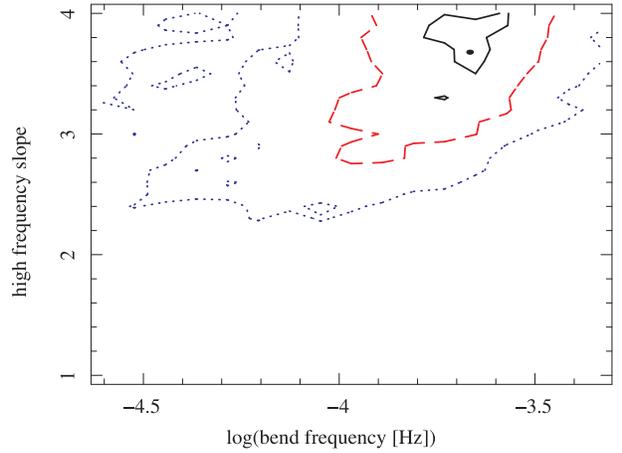,angle=270,width=8.0cm}
 \caption{\label{con1-3}Confidence contours of the bending power law fit to the 1--3 keV PDS. The 68\%, 90\% and 99\% levels are shown by the solid, dashed and dotted lines, respectively. The best fitting values are marked by the black dot. In this energy band and also in 3--10 keV, the lower 90\% and 99\% confidence limits on the break frequency are contained in the parameter space explored, i.e. at frequencies higher than the lowest bin frequency which is determined by the length of the observation.} 
\end{figure}

\subsection{Bend frequency vs H$_\beta$ relation}
\begin{figure}
\psfig{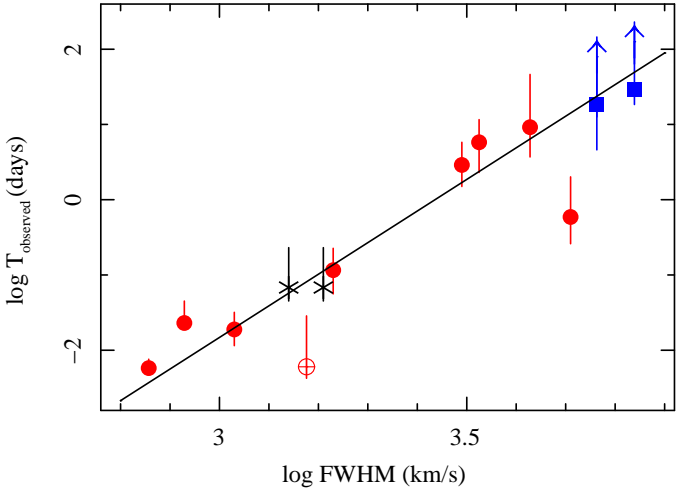}
\caption{\label{hb} H$_{\beta}$ line width vs break time-scale: the solid black line represents the best-fitting relation from \citet{mchardynat} who used the 9 AGN plotted as red circles and the low luminosity AGN NGC4395 plotted as an open crossed circle. The 2 AGN (green filled squares) whose upper limit on the break time-scale is not constrained are plotted, but were not included in the fit. The black crosses represent the position of \mkn\ on this plot, using the best-fitting break time-scale of the 0.2-10 keV PDS and the two line width measurements of \citet{peterson}. }
\end{figure}

\citet{mchardynat} have shown for a sample of AGN with $f_b$ measured as above, that $f_b$ is well correlated to the width of the H$_\beta$ line in the optical spectrum, following the relation 
\begin{equation}
\label{eqlinewidth}
\log{T_b}=4.2 \log{H_{\beta,\rm FWHM}}-14.43,
\end{equation}
where $T_b$ is the break time-scale in units of days and $H_{\beta,\rm
  FWHM}$ is in units of km/s. \citet{peterson} used data from two
observational campaigns to measure $H_{\beta,\rm FWHM}$ of \mkn\
obtaining values of 1629$\pm145$ and 1375$\pm357$ km/s. According to
Eq. \ref{eqlinewidth}, these line-widths predict a break time-scale of $0.078 < T_b <
0.16$ days in the first case and $0.016 < T_b < 0.15$ days in the
second. The measured bend frequency corresponds to a time-scale of
$0.068 ^{+0.16}_{-0.023}$ days, consistent with both line width
measurements, within the uncertainties. Figure \ref{hb} shows the
position on \mkn\ on the line-width vs bend time-scale plot of
\citet{mchardynat}, for both line-width measurements.

\begin{table}
\begin{tabular}{llll}
Energy band [keV]& $f_b [10^{-4}$Hz] & $\alpha_h$&fit probability\\
\hline
\hline
0.2--10&1.7$_{-1.2}^{+0.85}$ &3.8$^{+**}_{-1.0}$&87.2\%\\
\hline
0.2--0.6&1.11$^{+0.5}_{-0.71}$ &3.6$^{+**}_{-1.0}$& 66.9\%\\
0.6--1&1.55$^{+1.12}_{-0.93}$&4.0$^{+**}_{-1.1}$&90.6\%\\
1--3&2.22$^{+0.45}_{-0.36}$&3.7$^{+0.3}_{-0.4}$& 41.1\%\\
3--10&2.38$^{+1.6}_{-0.41}$&4.0$^{+**}_{-0.2}$&34.3\%\\
\end{tabular}
\caption{\label{psresp} PSRESP fits to the PDS in different energy bands using a bending power law model with fixed low frequency slope $\alpha_l=1$. The errors correspond to 66\% confidence contours in the $\alpha_H$ vs $f_b$ plane, asterisks denote unbounded contours.}
\end{table}

\subsection{Energy dependence of the PDS}

\begin{figure}
\psfig{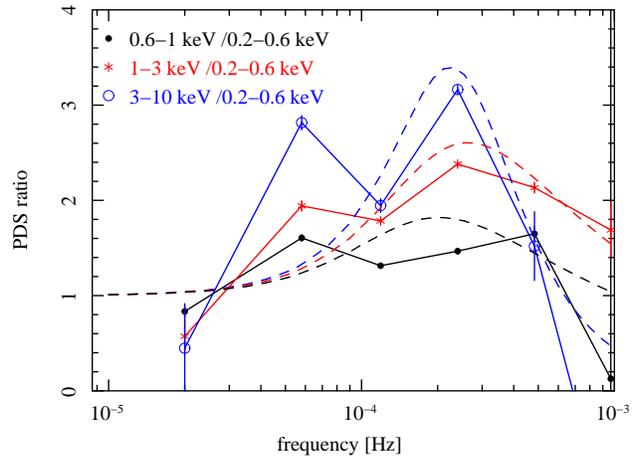}
 \caption{\label{psd_bands} Additional high-frequency power at high energies: Power spectra in the 0.6--1 (black), 1--3 (red) and 3--10 keV (blue) bands divided by the PDS of the 0.2-0.6 keV band. Higher energy bands show increasingly more variability power above $\sim 2\times 10^{-5}$Hz.}
\end{figure}

The PDS of \mkn\ shows more high frequency power at higher energies,
as is normally observed in AGN
\citep[e.g.][]{McHardy4051,markowitz,mchardy07}. Fitting the 0.2--0.6,
0.6--1, 1--3 and 3--10 keV PDS with PSRESP, keeping the low slope
fixed to 1, results in higher bend frequencies for higher energy bands
as shown in Table~\ref{psresp}, although they are partly within the
error from each other. The errors on these parameters are large,
however, because they take into account the fact that the observation
is only one realisation of a large set of possible observed PDS with
the same underlying power spectrum. Considering that the observations
in the different energy bands were made simultaneously, the
uncertainty in the \emph{relative} PDS shape is much smaller. To
illustrate the energy dependence more clearly, we calculated the
ratios of the harder band PDSs to the 0.2--0.6 keV PDS, which is shown
in Fig.~\ref{psd_bands}.  If the energy bands were completely
coherent, the only source of scatter would be the Poisson noise, which
adds different amounts of power to the PDS points of each light curve,
while the stochastic deviations of the coherent variations cancel
out. The variations in different bands are not completely coherent, as
shown below, so we considered the reduced coherence when calculating
the error bars on the PDS ratio. We assumed that 10\%, 20\% and 40\%
of the fluctuations in the 0.6--1, 1--3 and 3--10 keV light curves is
incoherent with the 0.2--0.6 keV fluctuations and added the
uncertainty in this incoherent part of the power to the uncertainty in
the power of the Poisson noise. These levels of incoherent
fluctuations were chosen to match the average coherence functions
between the corresponding energy bands, at least at low frequencies
where the coherence can be well measured, as described in
Sec.~\ref{coherence} below. With these error bars, the PDS ratios are
not consistent with a simple change in high frequency slope to account
for the additional high-frequency power, as in that case the ratios
would increase as a power law instead of leveling-off and
dropping. Equal power law slopes and different bend frequencies
produce PDS ratios of a stepped shape, that rise rapidly at
frequencies above the lower bend frequency and level-off above the
higher bend frequency. To produce a drop at even higher frequencies
would require a steeper high-frequency slope for the hard band.

For visualization purposes, we over-plotted in Fig.~\ref{psd_bands} the
ratios between bending power law model PDSs, with low frequency slope
$\alpha_l=1$ and different bend frequencies and $\alpha_h$. For the
model lines shown, we used the best fitting $f_b$ from Table
\ref{psresp} for the corresponding energy bands and fixed $\alpha_h=3$ for the
soft band. We allowed $\alpha_h$ of the hard bands to vary in order to
fit the amplitude of the PDS ratios of the data. The best fitting
values of $\alpha_h$ were 3.5 for 0.6--1 keV, 3.6 for 1--3 keV and 4.8
for 3--10 keV. The shape of the PDS ratio indicates that higher energy
bands have both a higher bend frequency and a \emph{steeper} high
frequency slope. This behaviour could indicate the presence of a
band-limited high frequency variability component with a hard energy
spectrum, adding on to the soft band PDS. This energy dependence, however, can also be produced by extended emitting regions filtering out the high frequency variability power. If higher energies are emitted by a more compact region, e.g. if the emitted spectrum hardens towards the centre, their PDSs will have higher bend frequencies than softer bands, while the high frequency slope can be equal or steeper (see figures 2 and 6 in \citealt{arevalo}).

\section{Coherence}
\label{coherence}
We will now investigate the correlation between variations in different energy bands. The coherence $\gamma^2$, i.e.\ the degree of linear correlation
between two light curves, can be computed as a function of Fourier
frequency using the cross spectrum $C(f)=S^*(f)H(f)$, where $S(f)$ and
$H(f)$ are the Fourier transforms of the soft and hard band light
curves $s(t)$ and $h(t)$, respectively.
 
\begin{equation}
\gamma ^2 (f_{i})=\frac{ \langle {\rm Re}_C (f_{i}) \rangle ^2 +
\langle {\rm Im}_C (f_{i}) \rangle ^2}{\langle | {\rm S}(f_{i})|^2
\rangle \langle | {\rm H}(f_{i})|^2\rangle }
\end{equation} 
where ${\rm Re}_C (f_{i})$ and ${\rm Im}_C (f_{i})$ are the real and
imaginary parts of the cross spectrum $C(f)$ and angle brackets
represent averaging over independent measurements. In our case, the
independent measurements are taken from consecutive Fourier frequencies
in a frequency bin. Coherence values range from 1, for totally
coherent , to 0 for totally incoherent variations. See
\citet{Vaughan_coh} for a detailed description of the coherence
function interpretation and error calculation.

\begin{figure*}
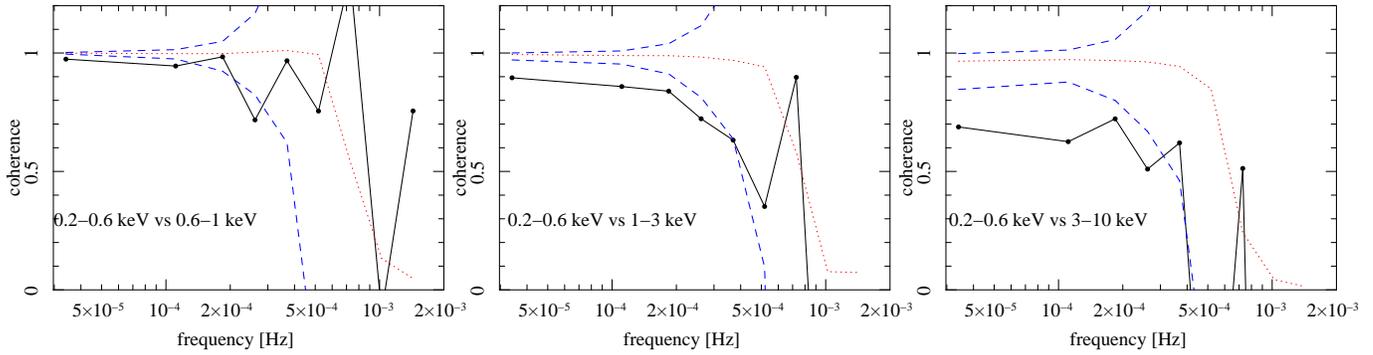

\begin{minipage}[b]{.33\textwidth}
\centerline{\psfig{file=arevalo_fig6.ps,angle=270,width=6.5cm}}
\end{minipage}
\begin{minipage}[b]{.33\textwidth}
\centerline{\psfig{file=arevalo_fig7.ps,angle=270,width=6.5cm}}
\end{minipage}
\begin{minipage}[b]{.33\textwidth}
\centerline{\psfig{file=arevalo_fig8.ps,angle=270,width=6.5cm}}
\end{minipage}
\caption{\label{coh}Coherence as a function of Fourier frequency between 0.2--0.4 and 0.4--0.6 keV energy bands (left), 0.2--0.6 and 1--3 keV bands (centre) and 0.2--0.6 and 3--10 keV bands (right), plotted in solid black lines. The dot-dashed red lines represent the median of a distribution of simulated coherence functions, constructed from intrinsically coherent light curves affected only by Poisson noise. For the simulations we used the count rate and underlying PDS shape of the corresponding light curves. The dashed blue lines represent the 95\% upper and lower limits of this distribution, marking the scatter expected in the case of perfectly coherent light curves, distorted by Poisson noise and the sampling pattern in the same way as the real data. Evidently, the measured coherence is significantly lower than 1 between the 0.2--0.4 and 1--3 keV bands and between the 0.2--0.4 and 3--10 keV bands, even at the lowest frequencies sampled.}
\end{figure*}

Coherence functions were calculated between 0.2--0.6 keV and each of
the three harder bands (0.6--1, 1--3 and 3--10 keV) and the results
are shown in Fig.~\ref{coh}. A minimum of 8 points was included in
each frequency bin and the bin separation was fixed at 1.5 $\times
f_i$.

As the Poisson noise and other distorting effects in the observed
light curves can artificially decrease the coherence, we applied the
Poisson noise correction factor described in \citet{Vaughan_coh} and
estimated any additional deviations using simulated light curves. For
each pair of energy bands, we generated 1000 pairs of coherent
red-noise light curves using the method of \citet{Timmer}, with the
underlying PDS parameters obtained from the fits to the corresponding
real light curves. The average count rates of the simulated light
curves were made to match the measured values and Poisson deviates
were added to each simulated light curve point. The coherence was then
calculated for each simulated pair using the same sampling and Poisson
noise correction method as used for the real data. The mean of the
distribution of simulated coherence measurements for each frequency
bin is plotted as a red solid line in Fig.~\ref{coh} and the top and
bottom 95\% extremes of this distribution are plotted as dashed blue
lines in the same plot.
  
This analysis shows that the variations in the 0.2--0.6 keV and 0.6--1
keV energy bands are highly coherent ($\gamma \sim 0.95$) up to
frequencies of at least $2\times 10^{-4}$ Hz. Above this frequency
Poisson noise makes the coherence estimates unreliable so it is not
possible to detect a real drop in coherence. At larger separations
between the energy bands the coherence drops significantly, even at
the lowest frequency probed, $4 \times 10^{-5}$Hz, to $\sim 0.87$
between 0.2--0.6 and 1--3 keV and to $\sim 0.6$ between 0.2--0.6 and
3--10 keV light curves.
 
\section{Time lags}
\label{lags}
Time delays between variations in different energy bands can be
measured using the phase lag spectrum $\phi (f)$. This spectrum gives
the difference in phase between the Fourier transform components of
the soft and hard light curves, as a function of Fourier frequency $f$,
and it can be converted into a time-lag spectrum as $\tau
(f)=\phi(f)/(2 \pi f)$. See \citet{nowak_lags} for a full description of the lag spectrum measurement. In what follows we will refer to the time lag
spectrum simply as lag spectrum.

In all AGN where the lag spectrum has been measured, the hard bands lag
softer ones \citep{papadakis_lags,vaughanmcg,McHardy4051,arevaloark,markowitz} and the size of the lag increases with decreasing
Fourier frequency. In the
BHXRB Cyg~X-1 in the high/soft state, the lag spectrum has a power law dependence on frequency
with a slope of $\sim 0.7$ and a cut-off at high frequencies, close to
the break in the PDS. In the low/hard state, the lag spectrum shows
approximately the same power law dependence but with additional
structure in the form of steps. AGN data are normally not good enough
to detect such structure and only simple power laws with fixed slope
are fitted. A change in slope in the lag spectrum of an AGN has only
been detected significantly in Ark~564 \citep{arevaloark} but in that
case the `step' was too deep to resemble the lag spectrum of Cyg~X-1
in the low/hard state and rather resembles a Very High state.

\subsection{Lag spectrum}
\begin{figure}
\psfig{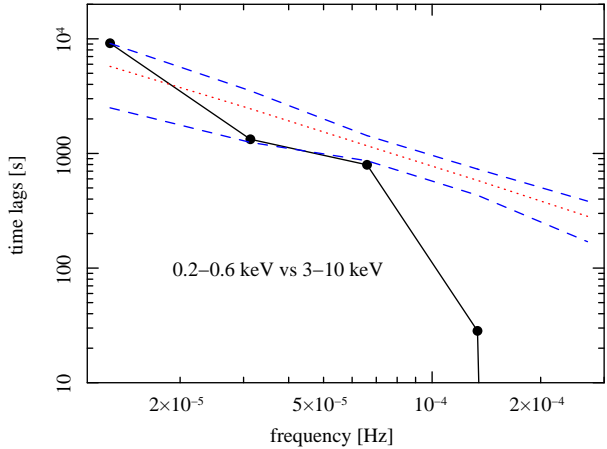}
 \caption{\label{lags3-10} Time lag as a function of Fourier frequency $f$ between the 0.2--0.4 keV and 3--10 keV energy bands over the frequency range where the coherence is well measured, the data points are shown in black markers joined by the solid line. The dotted and dashed lines represent the mean and top and bottom 95\% limits of a distribution of simulated lag spectra with an underlying $lag=0.082 f^{-1}$ lag spectrum (see text). The drop of the lags at high frequencies is greater than expected for this underlying model, showing that the real lag spectrum has either a slope steeper than -1 or it cuts-off at high frequencies. }
\end{figure}

The lag spectrum of \mkn\ between 0.2--0.6 and 3--10 keV energy bands
is shown in Fig. \ref{lags3-10}, where positive values indicate hard
band \emph{lagging}. A minimum of 2 points per bin was used and the
phase-lag measurements were averaged over frequency bins of width
2$\times f_i$, before being converted into time-lags. Notice that
the lags are shown only for frequencies below $3\times 10^{-4}$Hz,
where the coherence is still acceptably high.

As seen in other AGN and XRB, the hard band in \mkn\ lags the softer
band and the lag size decreases with increasing frequency with an
approximately power law dependence. The slope, however, appears
steeper than the $\tau(f)\propto f^{-1}$ model normally fitted to AGN
lag spectra. 

We used simulated light curves to determine whether the steepness of
the slope is produced by observational biases. Light curve pairs were
generated as in Sec. \ref{coherence}, introducing time lags between
the simulated soft and hard light curves of the form $\tau(f)=0.082
f^{-1}$, where the amplitude of the lag spectrum was obtained from a
fit to the real data. The simulated light curves were sampled
following the pattern of the real data and lag spectra were calculated
for each pair using the same method and binning. The mean of the
distribution of 1000 simulated pairs is shown by the red dotted line
in Fig.~\ref{lags3-10} and the top and bottom 95\% extremes of the
distribution are plotted in blue dashed lines. The real lag spectrum,
plotted in black markers joined by a solid line in the same figure,
drops more steeply than the simulations, indicating that the drop in
high-frequency lags is not due to observational biases on an
intrinsically $1/f$ lag spectrum. We repeated the same experiment
using the best-fitting power law model of the real data,
$\tau(f)=8.8\times 10^{-6}f^{-1.84}$, as the underlying lag model for
the simulations. In this case, the real data falls within the 95\%
extremes of the distribution of simulated lag spectra, but this model
over predicts the scatter, putting almost the entire 95\% lower limit
on negative lag values. Fitting the lag spectrum with a range of
slopes between -1 and -1.84 and repeating this experiment produced
simulated lag distributions that are either too tight and do not
include the high-frequency lag points and/or too wide and over predict the
scatter in the measured lag spectrum at low frequencies. Therefore, the lag spectrum is
not consistent with a single power law and the bend down or cut-off at
around $10^{-4}$ Hz is real. \citet{arevaloark} found a similar change
in slope in the lag spectrum of Ark~564, in that case however, the
bend was found a decade in frequency below the bend frequency of the
PDS, while in \mkn\ the bend in the PDS and lag spectrum appear at the
same frequency, within the uncertainties.

\subsection{Energy dependence of the time-lags}

The length of the time lags increases with the separation between the
energies of the bands considered. We calculated lag spectra of the
0.2--0.4 keV to the 0.4--0.6, 0.6--1, 1--2, 2--5 and 5--10 keV bands,
finding that the higher the energy band, the more it lags the 0.2--0.4
keV band. The fractional lags, $\tau$/time-scale, are approximately 0.72 \% for
  0.4--0.6 keV, 1.5\% for 0.6--1 keV, 4.2\% for 1--2 keV, 6.2\% for
  2--5 keV and 12\% for 5--10 keV. Each lag spectrum resembles the one
  shown in Fig. \ref{lags3-10}, having a cut-off at high frequencies. These fractional lag values only represent the low-frequency part
  of the spectra.

\begin{figure}
\psfig{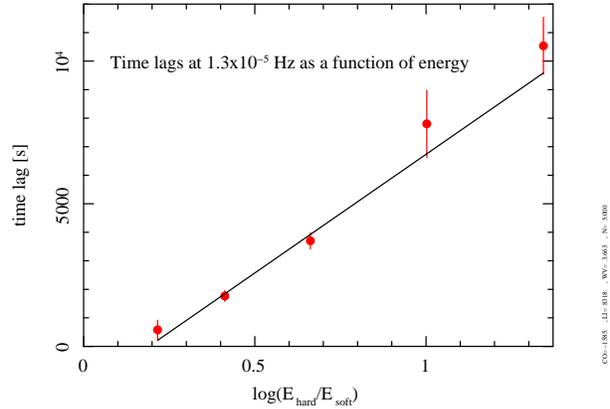}
\caption{\label{lagsvsE} Time lag at frequency $1.3\times10^{-5}$ Hz as a function of energy. The logarithm of the ratio between the average energies of the each pair of energy band is plotted in the x axis. The solid line represents a linear fit to the data plotted.  }
\end{figure}

In Fig.~\ref{lagsvsE} we plotted the lowest-frequency lag for each
pair as a function of the ratio between the average energy of the
bands compared. The lags increase in an almost log-linear manner with
increasing energy ratio. The best fitting log-linear model, shown by
the solid line in the plot has a form
\begin{equation}
\tau=(8317\pm 1138) \log{\frac{E_{\rm hard}}{E_{\rm soft}}} - (1585\pm580) \quad \rm sec.
\end{equation}
For a perfect log-linear relation, the intercept should equal zero,
i.e. lags between identical energy bands should be 0. The intercept of
the fitted relation is significantly different from this value so the
behaviour of the time lags in \mkn\ is not entirely consistent with
Cyg~X-1 \citep{nowak_lags}. Notice however, that the log-linear relation in Cyg~X-1 was
calculated for energies above 2 keV and in \mkn\ we used a baseline
energy band of 0.2--0.4 keV. It is possible that other spectral
components in \mkn\ below 2 keV, in particular the soft excess might
be suppressing the intermediate-energy lags.

\subsection{Interpretation of the lag spectrum}

We applied the propagating fluctuation paradigm of \citet{lyubarskii}
to attempt to reproduce the observed lag spectra. In this scenario,
the fluctuations are produced by the accretion flow over a large
range in radius, and travel towards the centre to modulate the X-ray
emitting region. We assume that the fluctuations are produced and
propagated on the viscous time-scale of a thick accretion flow and
that the emitted energy spectrum hardens towards the centre, therefore
producing hard lags.

The travel time $tt$ of a fluctuation of frequency $f$, from its
radius of origin $r_f$ to any radius $r$ is given by
\begin{equation}\label{tt}
tt(r)=\int^{r_f}_{r} \frac{dr}{v(r)}=\frac{2}{3(H/R)^2 \alpha}(r_f^{3/2}-r^{3/2})
\end{equation}
where the second equality assumes a propagation speed $v(r)=(H/R)^2
\alpha r^{-1/2}$ in units of $c$ and $r$ in units of
$R_g=GM/c^2$. This speed is taken from the standard disc prescription \citep{shakura}, where $\alpha$ is the viscosity parameter and $H/R$ the disc thickness
to radius ratio.

The time lags arise from different radial emissivity profiles
$\epsilon(r)$ characterising the different energy bands. The weighted
travel time, $\bar\tau (f)$ of a fluctuation on frequency $f$ folded
through a given emissivity profile is
\begin{equation}
\bar\tau(f)=\frac{\int^{r_f}_{r_{in}}tt(r)\epsilon(r)2\pi r dr}{\int^{r_f}_{r_{in}}\epsilon(r) 2\pi r dr}
\end{equation}
We will replace the normalisation factor $\int^{r_f}_{r_{in}}\epsilon(r) r
dr$ by $E$ in what follows and cancel the $2\pi$ factors. Finally, the time lag between two energy
bands with emissivity profiles $\epsilon_a(r)$ and $\epsilon_b(r)$ is the
difference between the corresponding weighted travel times,
\begin{equation}
\tau(f)=\bar\tau_a(f)-\bar\tau_b(f) =\int^{r_f}_{r_{in}}\left[ tt(r)\frac{\epsilon_a(r)}{E_a}-tt(r)\frac{\epsilon_b(r)}{E_b}\right] r dr 
\end{equation}
replacing $tt(r)$ by the expression in Eq.~\ref{tt} and cancelling terms gives:
 \begin{equation}\label{eqlag}
\tau(f)=\frac{-2}{3(H/R)^2 \alpha}\int^{r_f}_{r_{in}} r^{5/2}\left(\frac{\epsilon_a(r)}{E_a}-\frac{\epsilon_b(r)}{E_b}\right) dr 
\end{equation}
Notice that the length of the time lag is not simply proportional to
$1/(H/R)^2\alpha$ because the integration limit $r_f$ is also a
function of these disc parameters. If the disc thickness or $\alpha$
decrease, the propagation speed decreases making the lag longer, but
the fluctuations on a given frequency are produced at a smaller radius
$r_f$, so they travel a shorter distance and so $\tau(f)$
remains at approximately the same level. 

Equation~\ref{eqlag} gives the lag spectrum between energy bands
characterised by emissivity profiles $\epsilon_a$ and
$\epsilon_b$. From standard disc theory, the total energy dissipated
per unit area follows a $\epsilon\propto r^{-3}$ relation. As a simple
example we used similar power law emissivity profiles but with higher
exponents for higher energy bands, to produce a spectrum that hardens
towards the centre.  

\citet{kelly} measured a mass for \mkn\ of $6.6^{+6.3}_{-3.2}\times
10^{6}M_\odot$, we used this mass estimate to calculate $R_g/c=33 sec$
to compare model predictions to the real data. In this model, both the
bend in the PDS and the cut-off in the lag spectrum are related to the
characteristic time-scale of the innermost region of the accretion
flow. This normally requires thick disc parameters to produce the
power observed at the highest frequencies. We used $(H/R)\alpha=0.1$
to put the viscous time-scale, $t_v=r/v_{\rm visc}(r)$, at a radius of
$6R_g$ (i.e.\ the last stable orbit of a non-rotating black hole) at
$10^4$ s for a $6.6\times 10^{6}M_\odot$ black hole, which is
appropriate for modelling \mkn .

\begin{figure}
\psfig{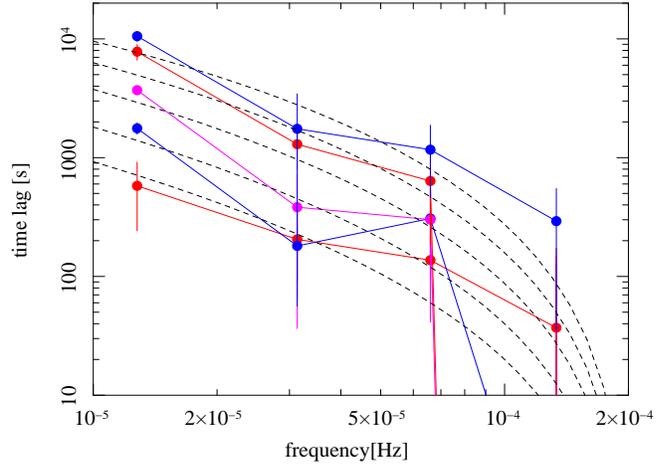}
\caption{\label{lags_bands} Time lag as a function of Fourier frequency between the 0.2-0.4 keV energy band and consecutively higher energy bands. The length of the lag increases with the separation of the energy bands. The dashed lines represent model fits as explained in Sec. 5.3   }
\end{figure}

The measured lag spectra between the 0.2--0.4 keV band and all harder
bands are shown by the markers and solid lines in
Fig.~\ref{lags_bands}, where the higher energy hard bands produce
longer lags. The quality of the data does not allow a detailed fit to
each lag spectrum but we attempted a simple implementation of the
model described above to make a broad comparison.  The dashed lines in
Fig. \ref{lags_bands} represent the model fits, where we assumed
parameters $(H/R)\alpha=0.1$, innermost radius of the disc $r_{\rm
  in}=6R_g$ and an emissivity profile for the 0.2--0.4 keV band of
$\epsilon\propto r^\beta$ with $\beta_{0.2-0.4}=-3.0$. The other
energy bands were also assumed to have power law emissivity profiles
and the exponents were allowed to vary in order to fit the data. The
standard viscous propagation speed resulted too high and produced lags
consistently shorter than the observed values. We reduced the
propagation speed by a factor of two for the fits, which reproduced
the observed average lags. The models shown in Fig.~\ref{lags_bands}
have exponents of -3.08, -3.15, -3.33, -3.6 and -3.8 for the 0.4--0.6,
0.6--1, 1--2, 2--5 and 5--10 keV bands, respectively.

The model does reproduce the power law shape of each lag spectrum and
predicts a cut-off at the right frequency. The range of lag spectra
amplitudes can be reproduced with moderate differences in the
emissivity profiles of the energy bands. These energy-dependent
emissivity profiles can be translated into a radial dependence of the
emitted spectrum. The exponents stated above produce spectral slope
changes from $\Delta \Gamma
=\Gamma(r_2)-\Gamma(r_1)=0.35\log(r_2/r_1)$ when we compare 0.2--0.4
to 0.4--0.6 keV energy bands up to $\Delta \Gamma =0.6\log(r_2/r_1)$
when comparing 0.2--0.4 to 5--10 keV energy bands, where $r_2>r_1$,
flux$(E)\propto E^{-\Gamma}$ and E is the average energy of the
band. This difference in spectral hardening implies that, for the
model to work exactly, either the radial emissivity profiles are not
exact power laws or that the emitted spectrum at each radius is not a
power law.

As \mkn\ is accreting at a high rate, an accretion disc with constant
scale height $H$, at least at the centre, can be more
appropriate. \citet{svensson} solve the structure equations of a thin
disc plus corona system, and show that the accretion flow represented
by the corona is radiation pressure dominated at accretion rates above
approximately 18\% the Eddington limit. As this is probably the case
in \mkn , if we identify the corona with the accretion flow producing
and propagating the variability, we should consider a scale-height
radial dependence of $H/R=r^{-1} \dot m (1-\sqrt{6/r})/2$. Assuming that
the same viscous velocity prescription applies in this case, then the
propagation velocity will contain an additional factor of $r^{-2}$
which reflects the change in $(H/R)^2$ with radius,
i.e. $v=(H/R)^2_o\alpha r^{-5/2}$, where $(H/R)^2_o$ is a fixed
value. The resulting travel time of the fluctuations changes to
$tt(r)=\frac{2}{7(H/R)_o^2\alpha}(r_f^{7/2}-r^{7/2})$, which can then
be used in Eq.~8 to calculate the lag spectra. The shape of the lag
spectra are approximately the same as in the case of constant
$(H/R)^2$, shown in Figure~\ref{lags_bands}. As long as the factor
$(H/R)^2\alpha$ is the same at the innermost radius, the constant
$H/R$ and constant $H$ models produce the bend at the same
frequency. The constant $H$ model however produces even shorter lags,
by approximately a factor of 4, so for this model to produce the
observed lag values the propagation speed should be reduced by a
factor of $\sim 8$ compared to the viscous velocity.

\section{Discussion and Conclusions}
\label{discussion}
We have studied the standard variability properties of \mkn\ using a
114 ks observation by \xmm . The main results are summarized below.

\begin{itemize}
\item The Power spectrum is well described by a bending power law model, with low frequency slope fixed at $\alpha_l=1$, best-fitting bend frequency of $f_b=1.7_{-1.2}^{+0.85} \times 10^{-4}$ Hz and high frequency slope of $\alpha_h=3.8^{+**}_{-1.0}$. 
\item The measured bend frequency coincides precisely with the value predicted by the $f_b$ vs H$_\beta$ line-width relation of \citet{mchardynat}, using the H$_\beta$ line-width measurements of \citet{peterson}, which demonstrates the validity of this relation.
\item The power spectrum is energy-dependent with more high-frequency power at higher energies. The difference in the PDSs of different energy bands indicates a shift in the PDS bend to higher frequencies for higher energy bands.
\item The degree of coherence between variations in different energy bands decreases with the energy separation of the bands considered. The coherence is significantly lower than 1, between the 0.2--0.6 and the 1--3 keV and 0.2--0.6 and 3--10 keV energy bands, even at the lowest frequencies sampled (4$\times 10^{-5}$ Hz).
\item Time lags are found between variations in different energy bands, where harder X-rays lag softer bands. The lag spectrum as a function of Fourier frequency can be fit with a single power law of slope $\sim -1.8$, much steeper than the lag spectra observed in other AGN. Alternatively, the lag spectrum of \mkn\ is consistent with a power law of slope $\sim -1$ dropping to a steeper slope at around $10^{-4}$Hz, where the bend in the PDS is also found. The latter shape resembles the high frequency end of lag spectra of Cyg~X-1 in both low/hard and high/soft states.  
\item The size of the time lags increases with energy separation following an almost log-linear relation. In all cases tested, using the energy bands 0.2--0.4, 0.4--1, 1--2, 2--5 and 5--10 keV, any band lags all softer ones, at least at low frequencies where the error bars are small. 

\citet{arevalo} have shown that the propagating-fluctuation paradigm of \citet{lyubarskii} reproduces the shape of the PDS and the rms-flux relation. This variability model is successful in producing variability over a very broad range of time-scales and it naturally predicts a cut-off at high frequencies. Complemented with an extended emitting region whose spectrum hardens towards the centre, this model can explain the time lags \citep{kotov01} and is compatible (but does not ensure) high coherence in the fluctuations of different energy bands. The extent of the emitting region in the model also acts as a low-pass filter reducing the high frequency power in the more extended energy bands, i.e. the softer bands, shifting the bend frequency to lower values. 

The energy dependence of the PDS, together with the shape of the lag spectra of \mkn\ are consistent with the predictions of this model, if the propagation speed is reduced by a factor of 2--8, compared to the viscous velocity, depending on the disc structure.  The lag spectra can be reproduced by a simple implementation of the model, with an emitted energy spectrum that hardens towards the centre. The energy dependence of the length of the time lags requires a maximum hardening of the spectrum with decreasing radius, between any radii $r_1$ and $r_2$, of $\Delta \Gamma =0.6\log(r_2/r_1)$. 

\end{itemize}

\section*{Acknowledgements}
We thank the referee for providing useful comments. This work has made use of observations obtained with \xmm , an ESA science mission with instruments and contributions directly funded by ESA member states and the US (NASA). PA and IMcH acknowledge support from STFC under rolling grant PP/D001013/1

\label{lastpage}
\end{document}